\documentclass[runningheads]{llncs}
\usepackage{graphicx}
\usepackage{hyperref}
\usepackage{booktabs}
\usepackage{multirow}
\usepackage{tabularx}
\newcommand{\tabularxmulticolumncentered}[3] 
    {\multicolumn{#1}
                 {>{\centering\hsize=\dimexpr#1\hsize+#1\tabcolsep+\arrayrulewidth\relax}#2}
                 {#3}}
\usepackage{amsfonts}
\usepackage{amsmath}
\usepackage{amssymb}
\usepackage{bbm}
\usepackage{smartref}
\usepackage{enumitem}
\begin{document}
%
\title{cOOpD: Reformulating COPD classification on chest CT scans as anomaly detection using contrastive representations}

\titlerunning{cOOpD}
%
%

\author{Silvia D. Almeida *\inst{1,2,3} \orcidID{0000-0002-4133-1194}\and
Carsten T. L\"uth *\inst{4,5}\and
Tobias Norajitra\inst{1,3}\and Tassilo Wald\inst{1,5} \and Marco Nolden\inst{1} \and Paul F. Jaeger\inst{4,5} \and Claus P. Heussel\inst{3,6}\and J\"urgen Biederer\inst{3,7}\and Oliver Weinheimer\inst{3,7}\and Klaus Maier-Hein\inst{1,3,5}}

\authorrunning{S. D. Almeida and C. T. L\"uth et al.}
%
\institute{
Division of Medical Image Computing, German Cancer Research Center, Heidelberg, Germany\\ 
\email{\{silvia.diasalmeida,carsten.lueth\}@dkfz-heidelberg.de} 
\and 
Medical Faculty, Heidelberg University, Heidelberg, Germany 
\and 
Translational Lung Research Center Heidelberg (TLRC), Member of the German Center for Lung Research (DZL), Heidelberg, Germany
\and
Interactive Machine Learning Group, German Cancer Research Center, Heidelberg, Germany 
\and 
Helmholtz Imaging, German Cancer Research Center, Heidelberg Germany
\and
Diagnostic and Interventional Radiology with Nuclear Medicine, Thoraxklinik at University Hospital, Heidelberg, Germany
\and 
Diagnostic and Interventional Radiology, University Hospital, Heidelberg, Germany
}
\maketitle              
\def\thefootnote{*}\footnotetext{These authors contributed equally to this work}\def\thefootnote{\arabic{footnote}}
\begin{abstract}

Classification of heterogeneous diseases is challenging due to their complexity, variability of symptoms and imaging findings.
Chronic Obstructive Pulmonary Disease (COPD) is a prime example, being underdiagnosed despite being the third leading cause of death. 
Its sparse, diffuse and heterogeneous appearance on computed tomography challenges supervised binary classification.
We reformulate COPD binary classification as an anomaly detection task, proposing cOOpD: heterogeneous pathological regions are detected as Out-of-Distribution (OOD) from normal homogeneous lung regions.
To this end, we learn representations of unlabeled lung regions employing a self-supervised contrastive pretext model, potentially capturing specific characteristics of diseased and healthy unlabeled regions.
A generative model then learns the distribution of healthy representations and identifies abnormalities (stemming from COPD) as deviations.
Patient-level scores are obtained by aggregating region OOD scores.
We show that cOOpD achieves the best performance on two public datasets, with an increase of 8.2\% and 7.7\% in terms of AUROC compared to the previous supervised state-of-the-art. 
Additionally, cOOpD yields well-interpretable spatial anomaly maps and patient-level scores which we show to be of additional value in identifying individuals in the early stage of progression.
Experiments in artificially designed real-world prevalence settings further support that anomaly detection is a powerful way of tackling COPD classification.

\end{abstract}
%
%
%
\section{Introduction}
\begin{figure}[h]
    \centering
    \includegraphics[width=1\textwidth]{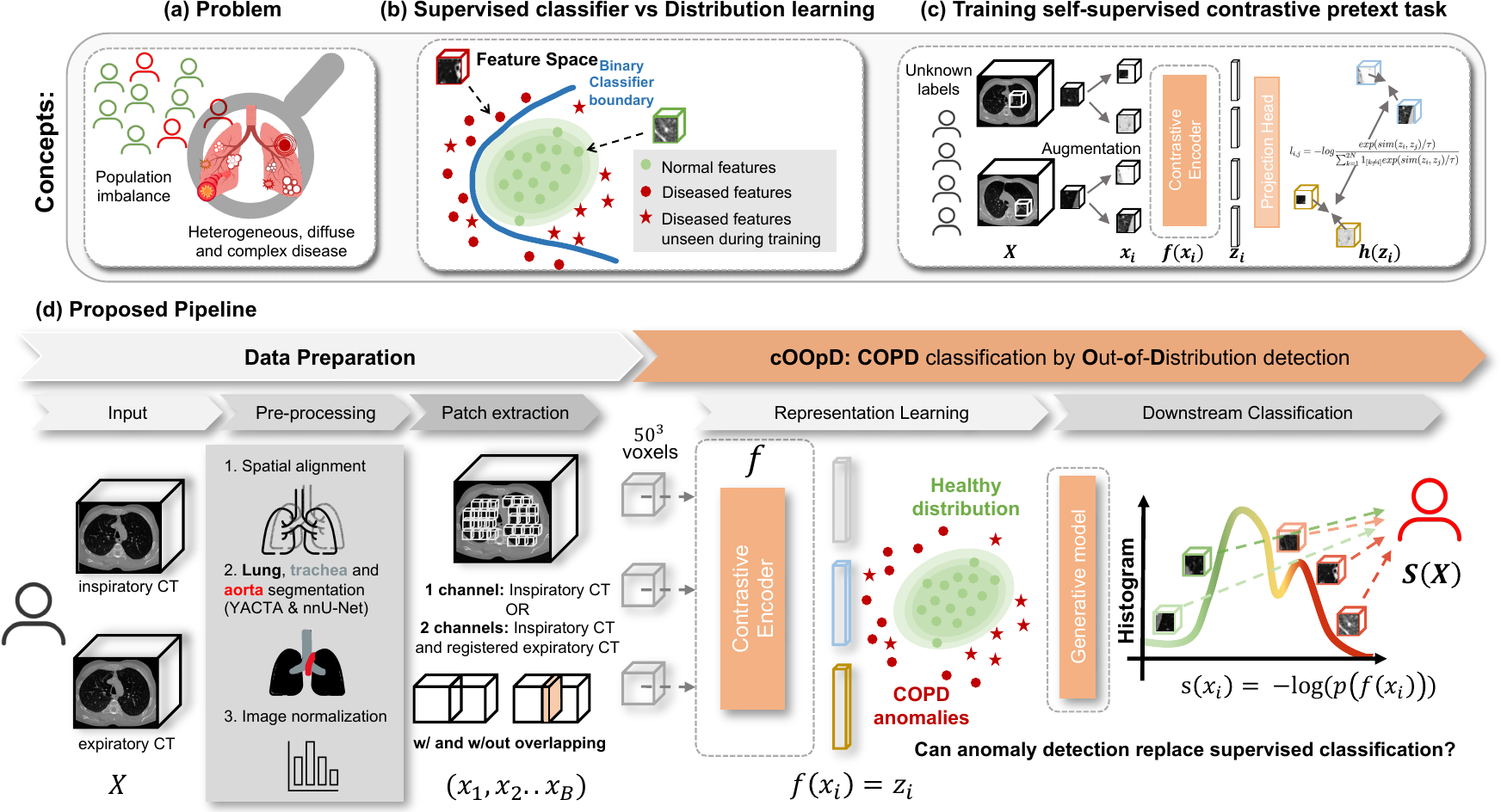}
    \caption{(a) Unbalanced prevalence of heterogeneous diseases in the population. (b) In the feature space, traditional classification methods may struggle with rare cases, while anomaly detection might lead to improved decision boundaries. (c) Self-supervised contrastive learning extracts meaningful representations from unlabeled lung patches, transitioning from voxels to features. (d) cOOpD during inference: paired CT scans are pre-processed, then patches are extracted and representations are obtained. Anomaly scores are assigned per patch, based on the distribution of healthy representations, which are then aggregated by patient.
    }
    \label{Figure:Method_1_2}
\end{figure}

By virtue of the human body's complexity, most diseases present phenotypic variability in terms of symptoms, rate of progression and imaging findings, which challenges diagnostic criteria.
Among many hard to diagnose diseases, Chronic Obstructive Pulmonary Disease (COPD) stands out, as it is extensively under- and misdiagnosed~\cite{soriano_screening_2009}, despite being the 3rd leading cause of death worldwide, with an estimated global prevalence of 10.3\%~\cite{adeloye_global_2022}.
Its pathological manifestations in the lung range from emphysema to airway disease, leading to a sparse, diffuse, and heterogeneous appearance, as shown in Fig.~\ref{Figure:Method_1_2}a.
Appropriate and earlier diagnosis that accounts for all of its manifestations is therefore of paramount importance for public health~\cite{gold_report2023}.\\
Considering the limitations of spirometry as the standard diagnostic method~\cite{gold_report2023}, computed tomography (CT) has emerged as a complementary tool for COPD characterization.
Initial efforts focused on identifying typical intensity and texture-level imaging features from either inspiration or expiration CT scans~\cite{bhatt_imaging_2019}.
With the advent of deep learning (DL), more complex supervised approaches have been proposed to tackle binary classification of COPD.
In this context, due to GPU memory constraints and large size of the images, different strategies to parcel a single 3D image as 2D slices~\cite{gonzalez_disease_2018,tang_towards_2020} or 3D patches~\cite{frangi_subject2vec_2018} have been pursued by supervised DL methods. 
Significant emphasis has been put on multiple instance learning (MIL) approaches~\cite{cheplygina_classification_2014,xu_dct-mil_2020,sun_detection_2022}, considering the spatial heterogeneity of COPD and that only a binary label is needed in case-finding scenarios.
Typically, for a supervised model to learn good decision boundaries, the labeled training dataset needs good coverage of the appearances of all classes.
However, good coverage of the diseased class can be difficult for low prevalence and heterogeneous diseases, making supervised models susceptible to fail on novel data points~\cite{kim_multicentre_2023}(Fig.~\ref{Figure:Method_1_2}b).
COPD fits exactly in this scenario, as its manifestations in the lung are diverse, in contrast to healthy individuals whose lungs are generally more uniform in appearance. 
This raises questions about the suitability of supervised models for COPD classification.\\
Instead of attempting to learn all possible complex manifestations of the disease, we ask: \emph{Could COPD be more accurately detected if considered as an anomaly from the distribution of healthy lungs?}\\
As previously reported for anomaly detection \cite{marimont2021anomaly,zhang2020hybrid}, modeling the distribution of normal samples in the latent space, instead of in the voxel space, has shown to be both feasible and desirable.
With this in mind, our contribution is two-fold:
\begin{enumerate}
    \item We show the benefit of reformulating COPD prediction as an anomaly detection task. Inspired by \cite{luth_cradl_2023}, we develop a generative model operating on the self-supervised representation space (Fig.~\ref{Figure:Method_1_2}c), learning the distribution of labeled healthy features and identifying unknown abnormal ones (stemming from \textbf{COPD}) as \textbf{O}ut-\textbf{o}f-\textbf{D}istribution (\textbf{cOOpD}). 
    cOOpD outperforms all compared DL-based supervised methods on two distinct public datasets, whilst maintaining performance in a scenario using a simulated real-world prevalence training dataset.
    \item We highlight the benefit of moving from voxels to representation space through a supervised method that leverages contrastive representations of lower dimensionality than voxels and outperforms voxel-based classifiers.
\end{enumerate}
\noindent
To the best of our knowledge, this work is the first to investigate anomaly detection in the context of a heterogeneous lung disease classification and has the potential to be applied to a wide range of diffuse diseases affecting large body areas.

\section{Method}
Our proposed method {cOOpD} aims at reformulating COPD classification as anomaly detection.
It is a self-supervised anomaly detection framework, inspired by the strategy of \cite{luth_cradl_2023}, optimized for diffuse lung diseases covering 3D multi-channel inputs, suitable augmentation strategies and patient-level aggregation of patch-level scores.
During inference (Fig.~\ref{Figure:Method_1_2}d), a sequence of $B$ 3D lung patches $\{x_i\}_{i=1}^B$ from a single or paired CT scan $X$ is extracted. 
Then, for each patch, a representation is obtained using a trained self-supervised contrastive encoder $z_i = f(x_i)$ (Sec. \ref{ss_contr}). 
Having learned the distribution of healthy patch-representations (Sec.~\ref{ss_gen-model}), the patch-level latent representation is given to a generative model $p(z)$, being attributed an anomaly score defined as the negative log-likelihood: $s(x_i)=-log(p(f(x_i)))$.
Several aggregation strategies $S(x)$ of these scores to patient-level were tested (Suppl.) including the mean, which was found to be the best performing and most conceptually meaningful strategy, as outlined in eq.~\ref{eq:score-mean-agg}.
\begin{equation}\label{eq:score-mean-agg}
    S(X) = \sum_{i=1}^B s(x_i)/B = - \log\left( \prod_{i=1}^B p\left(f(x_i)\right)^{1/B}\right)
\end{equation}
\subsection{Patch-Level Representations using Contrastive Learning}\label{ss_contr}
The latent representations of the encoder are learned with a self-supervised contrastive task, creating clusters based on semantic information.
For this, we follow the contrastive training described in~\cite{luth_cradl_2023} based on SimCLR~\cite{chen_simple_2020} with specific changes for medical images by providing more adequate mechanisms for 3D medical imaging.
Context and spatial information were covered by enabling 3D multi-channel patches as input, where each patch is then used as a singular sample for the contrastive task (Fig.~\ref{Figure:Method_1_2}c). 
Our augmentation strategy follows the approach of~\cite{shen_models_2019} with the following transformations:
Non-linear transformation based on the Bézier curve, local-pixel shuffling and in- and out-painting.
These were specifically designed for diffuse lung diseases and should force the encoder to learn patch representations capturing shape, texture, boundaries and context information. 
Preliminary experiments found that using all patches available per patient can introduce redundancy and substantially increase the computational cost (Suppl.). Therefore, a maximum of 100 patches per patient was set for training the self-supervised contrastive task. 
As an encoder, different 3D ResNet configurations (18 and 34) were tested.
\subsection{Generative Models operating on Representation Space}\label{ss_gen-model}
Once having extracted the latent representations, the distribution of normal representations is modeled, by fitting a generative model $p(z)$ on the representations of purely normal patches.
Patch normality is defined by $\%_{\mathrm{emphysema}}<1\%$ strictly applied to normal individuals, a very restrictive bound to guarantee that no intensity alterations could be present in the definition of normality. $\%_{\mathrm{emphysema}}$ is defined as the percentage of low attenuation areas less than a threshold of -950 Hounsfield units~\cite{bhatt_imaging_2019}. As generative models, Gaussian Mixture Model (GMM) and Normalizing Flow (NF) are employed.
While both are density estimation methods used to model $p(z)$, GMMs model the probability density function of the data as a weighted sum of Gaussian distributions, whereas our NF model uses the change of variable formula with a Gaussian prior.
The implementation of the NF is identical to \cite{luth_cradl_2023} consisting of fully connected affine coupling blocks and permutations based on the RealNVP architecture.
We fit several GMM with $\kappa \in {1,2,4,8}$ and a NF on representations of the encoder from the purely normal patches of healthy patients from the training dataset without any transformations. The best performing generative model is selected based on the validation set performance.
\section{Experiment Setup}

\subsubsection{Dataset \& Preprocessing:} 
Paired inspiratory and expiratory volumetric CT images were used from two nationwide multi-center studies (COPDGene~\cite{regan_genetic_2011}~\footnote{The COPDGene study (dbGaP \#28978) was funded by NIH grants U01HL089856 and U01 HL089897 and also supported by the COPD Foundation through contributions made by an Industry Advisory Board comprised of Pfizer, AstraZeneca, Boehringer Ingelheim, Novartis, and Sunovion.} and 
COSYCONET~\cite{karch_german_2016}), from which 5244 and 484 unique individuals were randomly selected, respectively (Suppl.). 
Binary classes were defined based on the Global Initiative for Chronic Obstructive Lung Disease (GOLD), a discrete score between 0–4.
The negative class (healthy) included never-smokers and individuals with a GOLD score of 0, while the positive class (diseased) included those with a GOLD score of 1 or higher.
This resulted in the prevalence of the positive class being 57\% for COPDGene and 85\% for COSYCONET. 
All trainings were performed on COPDGene which was split randomly into training (50\%), validation (25\%) and test (25\%) sets on the patient-level.
COSYCONET was entirely used as an external test dataset.
The data preparation process is illustrated in Fig.~\ref{Figure:Method_1_2}d and comprises the following sequential steps: 

\noindent \textit{Spatial alignment of paired inspiratory and expiratory CT images:}
Considering the potential of adding the expiratory scan as an extra channel as an indirect measure of gas trapping~\cite{bhatt_imaging_2019}, the paired images were geometrically aligned. Having the inspiratory image as the fixed image, an adaptation of~\cite{staring2010pulmonary} was performed. 
\noindent \textit{Lung parenchyma segmentation for patch extraction:}
Lung masks were generated on the inspiratory image space using a nnU-Net model~\cite{isensee_nnu-net_2021} on YACTA~\cite{achenbach_vollautomatische_2004} segmentation masks, a validated intensity-based method.

\noindent \textit{Intensity normalization:}
Inter-scanner variability was addressed by normalizing the intensity values to a scale between 0 (air) and 1 (tissue)~\cite{kim_improved_2014}. Mean intensity values for air and tissue were derived from segmented tracheal and aortic regions, respectively, obtained using a pre-trained nnU-Net model (Task\_055\_SegTHOR). Additionally, all images were resampled to an isotropic resolution of 0.5 mm.

\noindent \textit{Patch extraction:}
Volumetric patches (50\textsuperscript{3} voxels) containing $>70\%$ of the lung were extracted from the lung parenchyma of aligned inspiratory and expiratory CT images. The chosen size covered the secondary pulmonary lobule, the basic unit of lung structure~\cite{webb_thin-section_2006}. Two different patch overlapping strategies were implemented (0\% and 20\%) on inspiratory (1-channel) and inspiratory + registered expiratory (2-channels) images. Thus, four different configurations of input patches were tested.

\noindent \textbf{Baselines:}
State-of-the-art (SotA) baselines were applied to 2D slices and 3D patches.
A \emph{2D-CNN}~\cite{gonzalez_disease_2018} was employed at the patient-level. An end-to-end 3D patch classifier with score aggregation (\emph{PatClass}), an MIL approach with a Recurrent Neural Network as aggregation (\emph{MIL+RNN})~\cite{campanella_clinical-grade_2019} and an Attention-based MIL (\emph{MIL+Att}) (similar to~\cite{sun_detection_2022}, adapted from~\cite{ilse_attention-based_2018}) were employed at the patch-level. Implementation was performed as described in the original works, with adaptations to 3D, when required (Suppl.).

\noindent \textbf{Contrastive representations ablation:}
The contrastive latent representations' usefulness was evaluated with a supervised method (ReContrastive) that maps the latent representations back to their position in the original image, producing a 4D image, where the 4\textsuperscript{th} dimension is the length of the latent representation vector (Suppl). 
The produced image is then used as input for a CNN classifier. Training was performed for 500 epochs using the SGD Optimizer, a learning rate of 1e-2, Cosine Annealing~\cite{loshchilov_sgdr_2016} and a weight decay of 3e-5. A combination of random cropping, random scaling, random mirroring, rotations, and Gaussian blurring was employed as transformations.

\noindent \textbf{Evaluation metrics:}
We used Area Under Receiver Operator Curve (AUROC) and Area Under Precision Recall Curve (AUPRC) as the default multi-threshold metric for classification. 
AUROC is used as the main evaluation metric since it is less sensitive to class balance changes.

\noindent \textbf{Final method configurations:} These were chosen based on the highest AUROC on three experiment runs on the validation set.
The best patch extraction configuration for all tested 3D methods was two-channel (inspiratory and registered expiratory) with 20\% patch overlap.
The best performance was always achieved with a ResNet34.
For our proposed cOOpD method, GMM with $\kappa = 4$ was found to be the best performing generative model. 

\noindent \textbf{Real-world prevalence ablation:}
Given the global prevalence of COPD at 10.3\%~\cite{adeloye_global_2022}, we further evaluated the top two performing approaches in scenarios designed to approximate this real world prevalence. To better reflect these conditions, the diseased class in the COPDGene training set was undersampled to 5\%, 10.3\% and 15\% while keeping all samples from the normal class, limiting the diversity of the diseased class in the training set (instead of oversampling the normal class). 

\section{Results}
\begin{figure}[h!]
    \centering
    \includegraphics[width=1\textwidth]{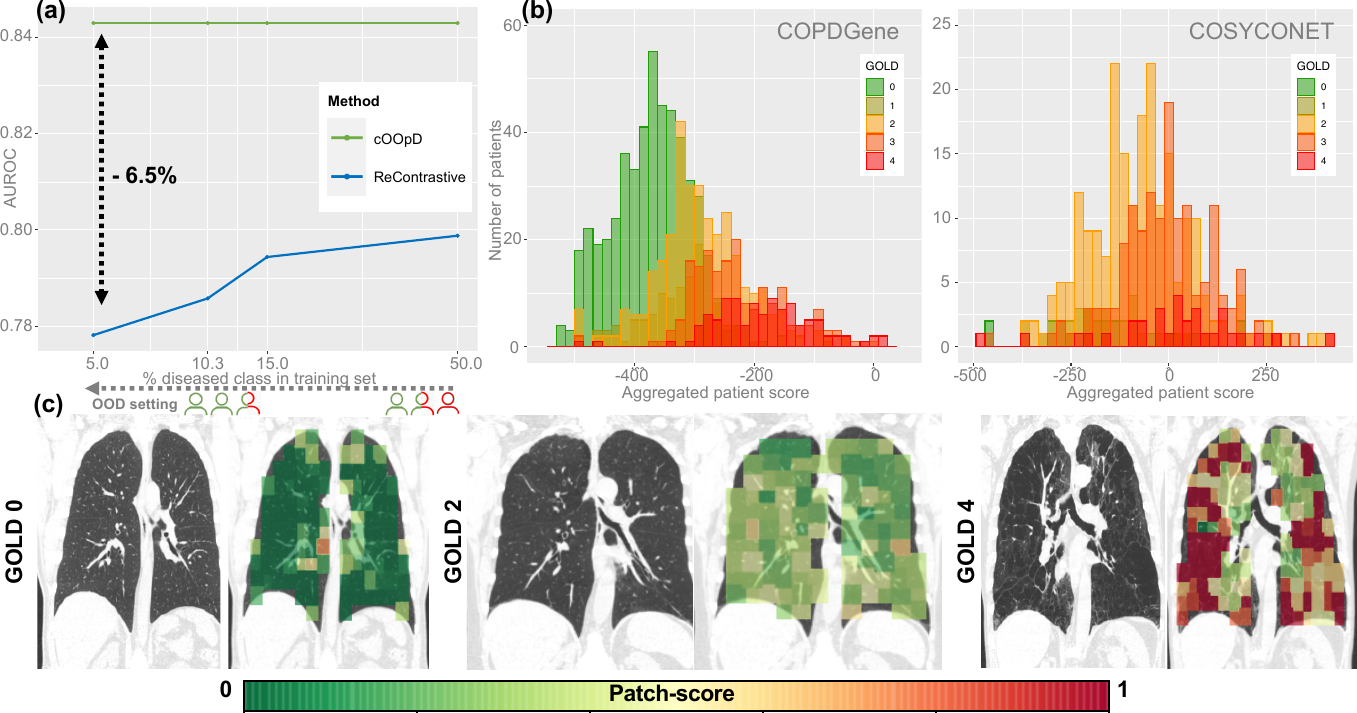}
    \caption{(a) Real world AUROC (5\%, 10.3\% and 15\% of diseased class prevalence) for the top two performing methods compared to the baseline (50\%). (b) Patient distribution by mean aggregated scores from cOOpD, colored by the function risk score (GOLD), for COPDGene and COSYCONET. (c) Coronal view of the cOOpD score map on three subjects with different degrees of severity. Min-max normalization was applied to patch scores, corresponding to the 5\textsuperscript{th} - 95\textsuperscript{th} percentiles of the internal testset.}
    \label{Figure:RealWorld}
\end{figure}

\begin{table}[]
\caption{Mean ± standard deviation in \% of 3 independent runs on the internal (COPDGene) and external (COSYCONET) test sets. 
Levels of statistical significance are denoted by (p$<$0.05\textsuperscript{*}/0.01\textsuperscript{**}) in comparison to the proposed method cOOpD (paired samples t-test).}\label{tab1}
\centering
\newcolumntype{Y}{>{\centering\arraybackslash}X}
\begin{tabularx}{\textwidth}{ llYYYY }
\toprule
\tabularxmulticolumncentered{1}{X}{\textbf{Input}} & \tabularxmulticolumncentered{1}{X}{\textbf{Methods}} & \tabularxmulticolumncentered{2}{X}{\textbf{COPDGene}} & \tabularxmulticolumncentered{2}{X}{\textbf{COSYCONET}}\\
                                &                                   &{AUROC} & {AUPRC} & {AUROC} & {AUPRC}       \\ 
\midrule
\multirow{1}*{2D image}   & \textit{2D-CNN}~\cite{gonzalez_disease_2018}                      & 55.6±2.5\textsuperscript{**}    & 72.0±1.5\textsuperscript{**}    & 57.0±8.0\textsuperscript{**}    & 84.6±1.4\textsuperscript{**}          \\ 
                                \hline
\multirow{5}{*}{3D patch}     & \textit{PatClass + RNN}   & 76.1±0.2\textsuperscript{**}    & 86.3±0.1\textsuperscript{**}    & 56.2±0.7\textsuperscript{**}   & 95.3±0.1\textsuperscript{*}          \\ 
                                & \textit{MIL + RNN}~\cite{campanella_clinical-grade_2019}                & 73.0±0.6\textsuperscript{**}    & 84.5±0.5\textsuperscript{**}    & 60.2±4.2\textsuperscript{*}    & 95.7±0.4\textsuperscript{*}          \\ 
                                & \textit{MIL + Att}~\cite{ilse_attention-based_2018,sun_detection_2022}          & 65.8±1.2\textsuperscript{**}    & 80.9±0.8\textsuperscript{**}    & 57.7±1.3\textsuperscript{**}    & 95.1±0.2\textsuperscript{*}         \\ 
                                & \textit{ReContrastive (ours)}     &   79.9±0.3\textsuperscript{**}	      &   88.5±0.2\textsuperscript{*}           &    53.3±0.1\textsuperscript{**}	& 95.0±0.1\textsuperscript{**}    \\ 
                                & \textit{cOOpD (ours)}             &  \textbf{84.3±0.3}             &   \textbf{89.7±0.2}            &    \textbf{67.9±0.7} &	\textbf{96.5±0.4}     \\ 
\bottomrule
\end{tabularx}
\end{table}
\noindent As shown in Tab.~\ref{tab1}, cOOpD outperforms all SotA supervised methods, achieving statistically significant improvements in terms of AUROC of 8.2\% compared to the best method on COPDGene (PatClass+RNN), and 7.7\% on COSYCONET (MIL+RNN). 
ReContrastive, as a supervised ablation for assessing the advantage of using representations, also outperformed all the other voxel-based supervised strategies on the internal test set, by an AUROC difference of 3.8\% but shows a large performance drop leading to the worst AUROC on the external test set.
In the real-world ablation, as seen in Fig.~\ref{Figure:RealWorld}a, the best performing supervised method (ReContrastive) performance decreases with the diseased class prevalence, reaching a drop of 6.5\% compared to cOOpD.

\section{Discussion}
%
\noindent \textbf{Final method configurations:} 
The best working method configurations reflect the following properties of the task:
Using both inspiratory and expiratory images provides information about pulmonary vascular alterations and airway wall thickness not visible on the inspiratory scan alone, as in line with~\cite{cao2021potential}.
Finer grained information is captured using overlapping patches, which tended to work better in conjunction with deeper encoders.
Regarding our proposed method cOOpD, we note the following: 
We hypothesize that the latent space's complexity level is low, being easily covered with a simple generative model. 
As for the aggregation strategy, considering the spatial distribution of COPD, it can happen that only a small part of the lung is diseased. As the negative-log-likelihood has a lower bound but not an upper bound, a single patch having a high score leads to a high overall score when using mean aggregation, which is the desired behavior.
\\
\noindent \textbf{Should COPD binary classification be formulated as anomaly detection?}\label{ss_anomaly}
cOOpD performance shows to be significantly superior compared to all tested methods, on the COPDGene (internal) and COSYCONET (independent external) test sets. 
The lower performance in the external test set was consistent with all other methods.
There are several potential explanations for this.
Besides being a highly imbalanced dataset, all patients in COSYCONET have a diagnosis of COPD and only 15\% are categorized into GOLD 0 due to normal lung function.
We hypothesize that these 15\% "healthy" individuals have early signs of disease that are not captured by voxel-based methods but are being encoded by the latent representations. 
Considering that cOOpD was trained only on healthy representations from the COPDGene dataset, whose normal class consisted of never-smokers and GOLD 0 subjects, it can still outperform all the other methods, since the unseen traits of the disease are seen as anomalies.
The advantage of solely modeling the healthy distribution is further highlighted by the real-world experiments (Fig.~\ref{Figure:RealWorld}a), where cOOpD performance remains unaffected, when compared to the supervised ablation (ReContrastive).
Identifying people at risk for disease worsening is paramount for COPD management. 
The anomaly score per patient fulfills this risk assessment need, by exhibiting a clear relation to the exact GOLD stage (Fig.~\ref{Figure:RealWorld}b), even though it was never explicitly given the GOLD stage as a multi-class label. 
Further, the lung region scores enable spatial localization of anomalies, giving interpretability to the method (Fig.~\ref{Figure:RealWorld}c).
These findings support our approach of reformulating COPD binary classification as an anomaly detection task.\\
\noindent\textbf{Are self-supervised patch-level latent representations advantageous to voxels?}
Both methods working on the representation space (cOOpD and ReContrastive) outperform all voxel-based baselines on the internal test set.
Although for ReContrastive this improvement is no longer seen for the external test set, being the worst performing method, the early signs of disease for the healthy class of COSYCONET are likely being encoded by the latent representations, as mentioned earlier.
We hypothesize that this performance drop stems from the problem of supervised models depicted in Fig.~\ref{Figure:Method_1_2}b.
Combined with the cOOpD findings, this still supports the hypothesis that patch-level latent representations provide meaningful information and reduce the complexity of the problem.
\section{Conclusion}
Our proposed reformulation of COPD binary classification into an anomaly detection task (cOOpD) demonstrated superior performance compared to SotA methods. 
Additionally, the advantage of using latent representations was demonstrated.
The cOOpD approach also demonstrated stability in performance when trained on datasets with simulated real-world class imbalance.
Future work should focus on further validation on larger and more diverse datasets, longitudinal evaluation, and exploring its application to other heterogeneous diseases where annotated diseased data is scarce and access to healthy data is abundant.


\newpage
\bibliographystyle{splncs04}
\bibliography{paper0206_bib}
\newpage
\end{document}


\title{Supplementary Material of cOOpD}
\titlerunning{cOOpD}
%
\author{}
\institute{}
\maketitle              
%
%
%
%
\let\thefootnote\relax\footnotetext{All experiments were implemented in Pytorch and Pytorch-Lightning, using a NVIDIA GeForce RTX 2080 Ti GPU.}
\begin{table}[h]
\caption{Demographic and clinical characteristics of the two datasets \emph{M(IQR)}, median, interquartile range; \emph{GOLD}, Global Initiative for Chronic Obstructive Lung Disease}\label{tab1}
\centering
\newcolumntype{Y}{>{\centering\arraybackslash}X}
\begin{tabularx}{\textwidth}{ l Y Y }
\toprule
\textbf{Demographic characteristics}& \tabularxmulticolumncentered{1}{X}{\textbf{COPDGene}}                                    & \tabularxmulticolumncentered{1}{X}{\textbf{COSYCONET}}                     \\ 
\midrule
\multirow{1}*{Age, Years, \emph{M} (IQR)}             & 63 (56 - 69)     & 66 (42 - 71)               \\ 
\multirow{1}*{Sex, \% male (n)}   & 54.1 (2836)    & 60.4 (278)\\ 
\multirow{1}*{BMI, mean (SD)}   & 28.5 (5.8)    & 26.6 (4.8) \\ 
\multirow{1}*{\%GOLD $\geq$ 1, \% (n)}   & 56.9 (2986)    & 85.2 (392)\\ 
\multirow{1}*{Number of patches per patient, \emph{M} (IQR)}   &     &\\ 
\multirow{1}*{0\% config}   & 319 (267 - 382)    & 308 (250 - 363)\\ 
\multirow{1}*{20\% config}   & 5622 (523 - 745)    & 635 (521 - 747)\\ 
\bottomrule
\end{tabularx}
\end{table}
%
%
\begin{figure}[h]
    \centering
    \includegraphics[width=0.7\textwidth]{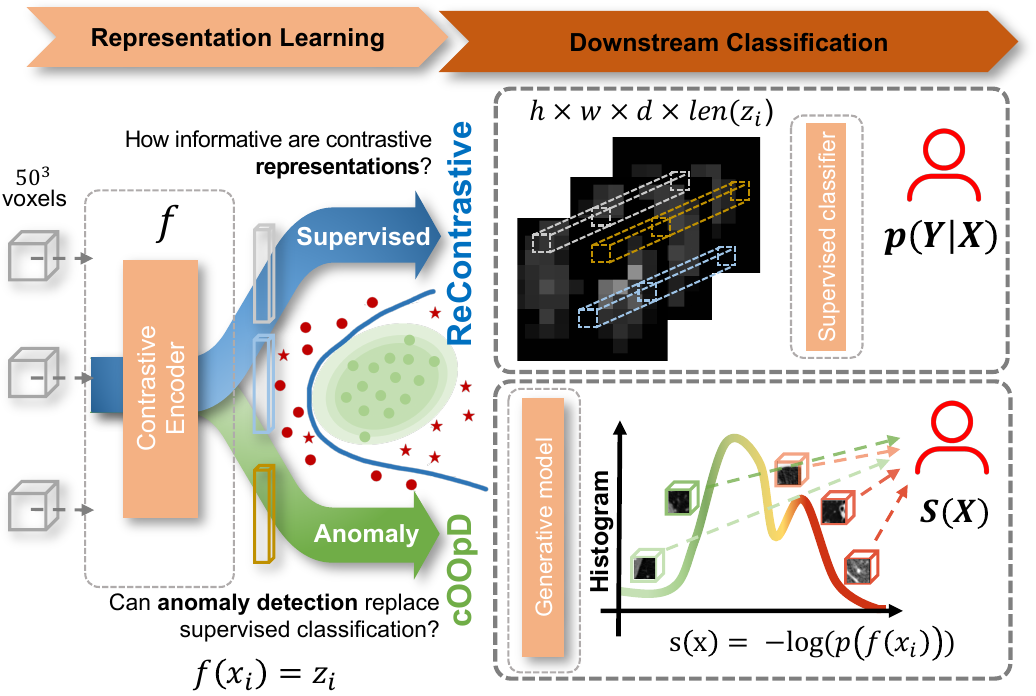}
    \caption{Pre-processed unlabeled patches are extracted and fed to a self-supervised contrastive task. Output representations value is address with ReContrastive, that maps the representations back to the original image and feeds it to a supervised classifier, whereas cOOpD allows for a reformulation of the task as anomaly detection.}
    \label{Figure:Method_1_2}
\end{figure}
%
\begin{table}[h]
\caption{Influence of setting a maximum number of patches per patient (PpP) (N=100, 200, maximum available) on training the contrastive task (Section 2.1), evaluated in the ReContrastive ablation. 
For comparable results, a maximum number of steps was set to 200000 and the validation data was set to fixed 300 patches. 
For simplicity, inspiratory images with 0\% patch overlap were used as input.
Our results show that the higher the number of patches, the higher the pretext task training loss. Regarding the evaluation on the downstream task, the highest performance is achieved for the experiment with the least amount of patches (N=100). \textit{ACC}: accuracy}\label{tab2}
\centering
\newcolumntype{Y}{>{\centering\arraybackslash}X}
\begin{tabularx}{\textwidth}{Y Y Y Y Y Y }
\toprule
\textbf{Max. \# of PpP} & \tabularxmulticolumncentered{1}{X}{\textbf{ACC}}                                    & \tabularxmulticolumncentered{1}{X}{\textbf{AUROC}}       & \tabularxmulticolumncentered{1}{X}{\textbf{Precision}}     & \tabularxmulticolumncentered{1}{X}{\textbf{Recall}}    & \textbf{Val loss pretext}                   \\ 
\midrule
\multirow{1}*{100}             & 79.7     & \textbf{79.5}     &    86.6     &        69.1      &        0.476 \\ 

\multirow{1}*{200}             & 78.0     & 77.8     &    89.5    &        62.8     &        0.512 \\ 
\multirow{1}*{all}             & 75.4     & 75.4     &    87.4    &        59.3     &        0.618 \\  

\bottomrule
\end{tabularx}
\end{table}
%
\begin{table}[h]
\scriptsize 
\caption{This end-to-end supervised patch binary classifier (PatClass) receives a 3D patch as input, which label corresponds to the patient label (0 or 1), and outputs a patch probability. Patient-level classification is obtained by aggregating the patch probabilities with a Recurrent Neural Network (RNN) as described in [5] which showed to be the best performing method on the validation set.}\label{tab3}
\centering
\newcolumntype{Y}{>{\centering\arraybackslash}X}
\begin{tabularx}{\textwidth}{ l Y }
\toprule
\textbf{Network \& Hyperparameters} & \tabularxmulticolumncentered{1}{X}{\textbf{Value}}                    \\ 
\midrule
\multirow{1}*{Architecture}             & 3D-ResNet34  \\ 

\multirow{1}*{Batch size}   & 64 \\ 
\multirow{1}*{N epochs}   & 100\\ 
\multirow{1}*{Optimization}   & Adam optimizer, learning rate 1e-4 and Cosine Annealing, weight decay 1e-5 \\ 
\multirow{1}*{Loss}   &   Cross Entropy Loss\\ 
\multirow{1}*{Transformations and probability}   & 5\%  elastic deformation, rotation, scaling, random crop, mirroring, gaussian noise and gaussian blur   \\ 
\bottomrule
\end{tabularx}
\end{table}
\begin{table}[h]
\caption{MIL + RNN implementation details. Implementation was as described in [5], except for the bellow hyperparameters. In our case, a slide corresponds to a patient and a tile to 3D patch.}\label{tab4}
\centering
\newcolumntype{Y}{>{\centering\arraybackslash}X}
\begin{tabularx}{\textwidth}{ l Y Y}
\toprule
\textbf{Network \& Hyperparameters}& \textbf{Batch size} & \textbf{N epochs}     \\ 
\midrule
\textbf{MIL} (3D-ResNet34)             & 480 & 200   \\
\textbf{RNN-10}             &  128 & 200 \\
\bottomrule
\end{tabularx}
\end{table}
\begin{table}[h]
\scriptsize 
\caption{Mean ± standard deviation of the AUROC in \% of 3 runs on the internal (COPDGene) validation set for the best configuration (ResNet34 with 20\% patch-overlap on the inspiratory + registered inspiratory). Best aggregation result is highlighted in bold and was then the only tested on the internal and external testsets.}\label{tab5}
\centering
\newcolumntype{Y}{>{\centering\arraybackslash}X}
\begin{tabularx}{\textwidth}{ YY YYYYYYYY}
\toprule
\textbf{Gen. Model} & \tabularxmulticolumncentered{8}{X}{\textbf{Aggregation}}                                \\ 
                                    &{Mean} & {Median} &{Q3} & {P95} &{P99} & {Max} &{Sum95} & {Sum99}      \\ 
\midrule
\textit{GMM 1}             & 85.8±0.0	& 85.1±0.0	& 85.4±0.0	& 85.7±0.0	& 85.8±0.0	&	85.0±0.0 & 	84.1±0.0	& 83.8±0.1                \\ 
\textit{GMM 2}                     & 86.0±0.5 &	85.2±0.3 &	86.0±0.1 &	84.7±0.7 &	79.5±3.7 &	71.8±10.1 &	80.4±3.9 &	75.4±7.9       \\ 

\textit{GMM 4}  & \textbf{86.2±0.5} &	84.4±0.9 &	85.8±0.8 &	85.5±0.7 &	83.3±2.0 &	72.8±5.0 &	84.1±1.8 &	79.6±4.5      \\ 
\textit{GMM 8}  & 85.9±0.4 &	83.5±2.0 &	84.0±0.9 &	84.2±0.7 &	81.3±1.4 &	71.0±1.6 &	82.8±1.1 &	77.8±1.9       \\ 
\textit{NF}   & 82.9±0.1 &	82.8±0.4 &	83.5±0.2 &	78.8±0.3 &	74.2±0.2 &	70.0±0.9 &	54.2±0.3 &	56.4±0.9       \\ 

\bottomrule
\end{tabularx}
\end{table}
\newpage